\documentclass{article}
\usepackage{spconf}
\usepackage[T1]{fontenc}
\usepackage{cite}
\usepackage{amsmath,amssymb,amsfonts}
\usepackage{algorithm}
\usepackage{algorithmic}
\usepackage{graphicx}
\usepackage{stfloats}
\usepackage{booktabs}
\usepackage{multirow}
\usepackage{textcomp}
\usepackage{xcolor}
\usepackage{hyperref}
\usepackage[capitalize]{cleveref}
\usepackage{mathtools}
\usepackage{bm}
\newcommand{\bx}{\bm{x}}            
\newcommand{\beps}{\bm{\varepsilon}} 
\begin{document}
\ninept

\title{ParaAnya: Accelerating Parallel Diffusion Sampling with Plug-and-Play Output Caching}

\name{
Chee-En Yu$^{1*}$,
Xiao-Xi Tan$^{1*}$,
Yi-Cheng Lin$^{1}$,
Yun-Shao Tsai$^{1}$,
Chee-An Yu$^{2}$,
Hung-yi Lee$^{1,3}$
}

\address{
$^{1}$National Taiwan University, Taiwan \\
$^{2}$University of Southern California, USA \\
$^{3}$Artificial Intelligence Center of Research Excellence, National Taiwan University, Taiwan \\
\thanks{\textsuperscript{*}Equal Contribution.}
}

\maketitle

\begin{abstract}
Diffusion models have achieved remarkable success in generative tasks, but their inherently sequential sampling process introduces a severe computational bottleneck. Recent Parallel-in-Time (PinT) solvers attempt to mitigate this by parallelizing generation across a sliding window of timesteps, advancing the window only when step-wise changes stabilize. However, this overlapping window mechanism forces the network to repeatedly evaluate the same timesteps. When the input variations between iterations are minimal, these redundant evaluations lead to significant computational waste.
To address this inefficiency, we propose ParaAnya, an output cache mechanism agnostic to the parallel sampling algorithm that can reduce the number of function evaluations (NFE).
ParaAnya caches input-output pairs of diffusion models and reuses the cached output at overlapping timesteps. By dispatching only cache-miss timesteps to GPU workers, our approach eliminates redundant computation while preserving the structure of the underlying algorithms' update rules.
We integrate ParaAnya into four representative parallel sampling algorithms and evaluate its performance on Stable Diffusion v1.5. Across four parallel samplers evaluated with DDIM on eight GPUs, ParaAnya provides $1.30$--$2.43\times$ speedups over their uncached counterparts and reduces NFE by up to 70.1\%, reaching up to a $5.62\times$ speedup over single-GPU serial sampling while maintaining comparable CLIP scores. Code is available at \url{https://github.com/XXIIIII/ParaAnya}
\end{abstract}

\begin{keywords}
Diffusion Models, Parallel-in-Time, Inference Acceleration
\end{keywords}

\section{Introduction}
\label{sec:background}
While diffusion models achieve state-of-the-art performance in image and audio generation\cite{ho2020denoising,dhariwal2021diffusion,rombach2022high,kong2020diffwave,liu2023audioldm}, their sampling process is inherently sequential, as each denoising step depends on the result of the previous timestep.
Moreover, generation quality is closely tied to the number of denoising steps, with more steps often yielding better samples at the cost of increased inference time\cite{nichol2021improved}. Consequently, the length of the denoising trajectory becomes a major bottleneck for diffusion sampling.
With multiple GPUs, conventional batch inference can improve \emph{throughput} by generating multiple samples in parallel, but it does not shorten the sequential denoising trajectory of an individual sample and therefore cannot reduce \emph{single-sample latency}.
To address this bottleneck, Parallel-in-Time (PinT) methods ~\cite{gander201550} expose parallelism along the temporal dimension of the diffusion trajectory, allowing multiple timesteps of a single sample to be processed concurrently. In this work, we aim to further improve the efficiency of PinT-based diffusion sampling without retraining the underlying diffusion model.
\begin{figure}[!t]
    \centering
    \includegraphics[width=\columnwidth, trim=0 0 0 0, clip]{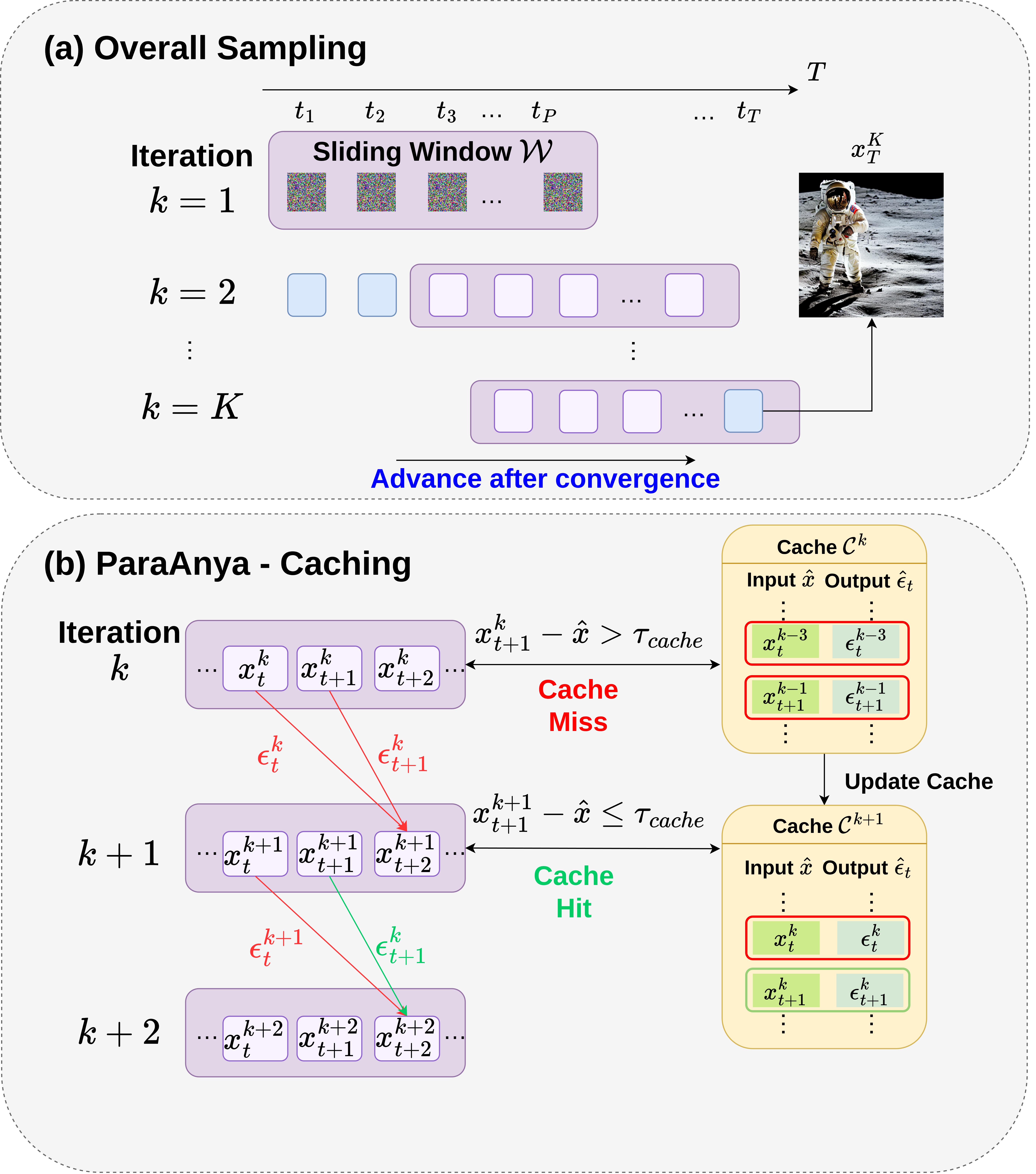}
    \vspace{-1em}
    \caption{(a) Overall Parallel sampling of Diffusion models with Picard iteration. Blue steps mean converged steps.
    (b) Caching mechanism of ParaAnya. Cache hits reuse the stored network output, while only cache misses are dispatched to GPU workers for batched evaluation. The solver then receives the complete set of drift values and proceeds with its original update rule.}
    \label{fig:method}
    \vspace{-1.5em}
\end{figure}
To overcome this inherently serial bottleneck, PinT methods parallelize diffusion sampling along the time dimension.
Existing approaches can be broadly categorized into Picard-based methods, including ParaDiGMS~\cite{shih2023parallel}, ParaTAA~\cite{tang2024accelerating}, and ParaSolver~\cite{lu2025parasolver}, and Parareal-based methods such as SRDS~\cite{selvam2024self}, which build on Picard iteration~\cite{ince1956ordinary} and Parareal~\cite{lions2001resolution}, respectively.
Despite their different formulations, these methods share a common principle: they first estimate the latent states at multiple timesteps in parallel and then iteratively refine these estimates across iterations. When the latent state at a timestep changes sufficiently little between consecutive iterations, that timestep is considered converged and its state can be finalized.
However, this iterative refinement introduces substantial redundant computation. Before convergence, the same timestep may be evaluated repeatedly across successive iterations, even when its input latent changes only slightly. Our preliminary study confirms that such cases occur frequently: profiling 100-step DDIM~\cite{song2020denoising} sampling over 30 prompts, we find that 39.93\% of all requested drift evaluations revisit a timestep with a mean latent change below 0.05. This observation motivates us to reuse previously computed outputs for similar latent inputs. Unlike prior caching methods such as DeepCache~\cite{ma2024deepcache}, which reuse information across adjacent denoising timesteps, our approach exploits redundancy across different PinT iterations at the same timestep.

We introduce ParaAnya, a caching mechanism that can be applied to the aforementioned PinT methods. ParaAnya stores an input-output pair for each timestep and directly reuses the cached output when the current input latent remains sufficiently close to the cached input. A new network evaluation is performed and the cache is updated only when the latent change exceeds a predefined threshold. Experiments on Stable Diffusion v1.5 demonstrate consistent improvements across four representative PinT samplers and all evaluated DDIM step counts, reducing network evaluations by up to 70.1\% and providing up to a $2.43\times$ additional speedup over their uncached counterparts. With eight GPUs, ParaAnya achieves up to a $5.62\times$ speedup over single-GPU sequential sampling while maintaining comparable CLIP scores~\cite{radford2021learning,hessel2021clipscore,ramesh2022hierarchical}, demonstrating the substantial performance gains enabled by caching within parallel diffusion sampling.

\section{Method}
\label{sec:method}

\subsection{Parallel Sampling Preliminaries}
Parallel-in-time (PinT) diffusion algorithms accelerate inference by evaluating the network at multiple timesteps concurrently.
Starting from an initial trajectory estimate, they iteratively refine the states at different timesteps until convergence. Picard-based methods, such as ParaDiGMS, ParaTAA, and ParaSolver, perform this refinement using a discretized Picard iteration:
\begin{align}
\bx_t^{k+1}
= \bx_0 + \frac{1}{T}
\sum_{i=0}^{t-1}s_{\theta}(\bx_i^{k}, i/T).
\label{eq}
\end{align}
where $k$ denotes the refinement iteration and $s_\theta(\bx,u)$ denotes the probability-flow ODE drift~\cite{song2020score}, which combines the current latent $\bx$ and the noise prediction $\beps_\theta(\bx,u)$ with coefficients determined by the noise schedule~\cite{lu2022dpm}.
Since all drift evaluations $s_{\theta}(\bx_i^{k}, i/T)$ depend only on states from the previous iteration, they can be evaluated in parallel across timesteps. Although individual methods employ different refinement strategies, they share an important characteristic: the network is repeatedly evaluated at the same timesteps as the trajectory is progressively refined.

In practice, GPU memory constraints often prevent the entire trajectory from being processed simultaneously. Parallel samplers therefore operate on a sliding window $\mathcal{W}$ containing $P$ active timesteps. As the refinement proceeds, the window advances once the leading states satisfy a convergence criterion, e.g.,
\begin{equation}
\frac{1}{D}||\bx_t^{k}-\bx_t^{k-1}||^2 \leq \tau,
\end{equation}
where $D$ is the dimensionality of the state and $\tau$ is a convergence threshold.

For Parareal-based algorithms such as SRDS~\cite{selvam2024self}, they have coarse solvers and fine solvers. The algorithm uses coarse solvers to run an initialization and uses fine solvers to get more fine-grained results.
Although their update structure differs from Picard iteration, they also revisit network evaluation points as trajectory estimates are refined.

Therefore, both Picard-based and Parareal-based parallel samplers repeatedly perform network evaluations at the same timesteps across refinement iterations. Moreover, as the trajectory converges, successive inputs to these evaluations can become increasingly similar. This repeated computation creates an opportunity to reuse previously computed network outputs through caching, thereby reducing redundant evaluations and improving inference efficiency.

\subsection{ParaAnya: An Output Caching for PinT diffusion solvers}
To eliminate redundant computation for latents that are near convergence, we introduce \textsc{ParaAnya}, an output cache placed before the drift evaluation. The coordinator maintains a cache $\mathcal{C}$ that associates each timestep $t$ with the most recently evaluated network input and its corresponding output:
\begin{align}
    \mathcal{C}: t \mapsto \left(\hat{\bx}_t,\, \hat{\beps}_t\right), \qquad \hat{\beps}_t = \beps_\theta\!\left(\hat{\bx}_t,\, t\right).
    \label{eq:cache}
\end{align}
Before evaluating the network, the coordinator measures how much the current latent $\bx_t^{k}$ has changed from the cached input $\hat{\bx}_t$ using the mean absolute deviation:
\begin{align}
    \delta_t^{k} = \frac{1}{D} \left\| \bx_t^{k} - \hat{\bx}_t \right\|_1,
    \label{eq:cache_delta}
\end{align}
where $\|\cdot\|_1$ is the entrywise $\ell_1$ norm. Dividing by the latent dimension $D$ makes $\delta_t^{k}$ a per-coordinate mean, allowing the cache threshold $\tau_{\mathrm{cache}}$ to remain comparable across different latent resolutions.
A timestep entering the window for the first time is treated as a cache miss. Otherwise, the current window $\mathcal{W}$ is partitioned into cache hits $\mathcal{H}$ and misses $\mathcal{M}$:
\begin{align}
    \mathcal{H}^{k} &= \left\{\, t \in \mathcal{W} : \delta_t^{k} < \tau_{\mathrm{cache}} \,\right\}, \nonumber \\
    \mathcal{M}^{k} &= \mathcal{W} \setminus \mathcal{H}^{k},
    \label{eq:hitmiss}
\end{align}
The drift output $\beps$ used by the solver is then assembled as
\begin{align}
    \beps_t^{k} =
    \begin{cases}
        \hat{\beps}_t & t \in \mathcal{H}^{k}, \\[2pt]
        \beps_\theta\!\left(\bx_t^{k},\, t\right) & t \in \mathcal{M}^{k},
    \end{cases}
    \label{eq:assemble}
\end{align}
Only the misses are dispatched to the workers, and all misses are evaluated together in a single batched forward pass.
After the evaluation, the cache is updated only for the misses:
\begin{align}
    \mathcal{C}[t] \leftarrow \left(\bx_t^{k},\, \beps_t^{k}\right), \qquad \forall\, t \in \mathcal{M}^{k},
    \label{eq:cache_update}
\end{align}
Therefore, whenever an entry $\mathcal{C}[t]=(\hat{\bx}_t,\hat{\beps}_t)$ exists, its stored output always satisfies
$\hat{\beps}_{t}=\beps_{\theta}(\hat{\bx}_t,t)$, preserving the cache invariant.
The cached or freshly computed network outputs $\beps_t^{k}$ are then used to compute the drift values at the current latent states. The sampler subsequently follows its original trajectory update, convergence check, and window advancement rules.

Parareal-based samplers such as SRDS also repeatedly evaluate the network within their coarse and fine solvers across refinement iterations. ParaAnya applies the same caching mechanism to these evaluations, reusing cached network outputs when the current latent is sufficiently close to the cached input at the same timestep. This reduces redundant network evaluations while leaving the Parareal update rule unchanged.

\begin{table*}[t]
\centering
\caption{Main results on COCO2017 using DDIM ($P\!=\!16$) on 8 NVIDIA Tesla V100-SXM2 32GB GPUs. ParaAnya variants achieve the fastest inference in all settings. (\textbf{bold}) while preserving quality ($\Delta\text{CLIP}$, 95\% CI). NFE is reported per generated sample.}
\label{tab:main}
\renewcommand{\arraystretch}{1.05}
\setlength{\tabcolsep}{3pt}
\small
\begin{tabular}{l l rrr rrr rrr}
\toprule
& & \multicolumn{9}{c}{DDIM} \\
\cmidrule(lr){3-11}
& & \multicolumn{3}{c}{$N=25$} & \multicolumn{3}{c}{$N=50$} & \multicolumn{3}{c}{$N=100$} \\
\cmidrule(lr){3-5}\cmidrule(lr){6-8}\cmidrule(lr){9-11}
Base method & Variant & NFE/img & $\Delta$CLIP & Speedup & NFE/img & $\Delta$CLIP & Speedup & NFE/img & $\Delta$CLIP & Speedup \\
\midrule
Serial & ---       & 25 & 31.24 & 1.00$\times$ & 50 & 31.25 & 1.00$\times$ & 100 & 31.26 & 1.00$\times$ \\
\midrule
\multirow{2}{*}{ParaDiGMS}
 & base   & 191 & 31.21 & 0.88$\times$ & 271 & 31.23 & 1.27$\times$ & 345 & 31.21 & 2.05$\times$ \\
 & ParaAnya & 79 & $+0.06\pm0.12$ & \textbf{1.80$\times$} & 81 & $+0.02\pm0.12$ & \textbf{3.09$\times$} & 120 & $+0.07\pm0.11$ & \textbf{4.12$\times$} \\
\midrule
\multirow{2}{*}{ParaSolver}
 & base   & 123 & 31.23 & 1.29$\times$ & 176 & 31.27 & 1.90$\times$ & 215 & 31.25 & 3.18$\times$ \\
 & ParaAnya & 68 & $+0.02\pm0.09$ & \textbf{2.27$\times$} & 70 & $+0.01\pm0.10$ & \textbf{3.71$\times$} & 104 & $-0.03\pm0.08$ & \textbf{4.98$\times$} \\
\midrule
\multirow{2}{*}{ParaTAA}
 & base   & 98 & 31.32 & 1.63$\times$ & 143 & 31.43 & 2.37$\times$ & 175 & 31.33 & 3.95$\times$ \\
 & ParaAnya & 72 & $-0.13\pm0.08$ & \textbf{2.36$\times$} & 75 & $-0.32\pm0.10$ & \textbf{3.96$\times$} & 101 & $+0.07\pm0.09$ & \textbf{5.62$\times$} \\
\midrule
\multirow{2}{*}{SRDS}
 & base   & 36 & 31.10 & 1.76$\times$ & 64 & 31.18 & 2.29$\times$ & 120 & 31.26 & 2.71$\times$ \\
 & ParaAnya & 25 & $-0.01\pm0.01$ & \textbf{2.66$\times$} & 50 & $-0.00\pm0.00$ & \textbf{3.18$\times$} & 100 & $-0.00\pm0.00$ & \textbf{3.53$\times$} \\
\bottomrule
\end{tabular}
\end{table*}

\section{Experimental Setup}

\subsection{Metrics}
\label{subsec:metrics}
We explore widely used metrics for our proposed ParaAnya: the number
of function evaluations (NFE), and CLIP score.

\subsection{Models and Testing Prompts}
We evaluate text-to-image generation with Stable Diffusion v1.5~\cite{rombach2022high} at $512\times512$ resolution, using a half-precision UNet with xFormers attention~\cite{lefaudeux2022xformers}.
All experiments use the same 1{,}000 prompts, obtained by shuffling the COCO 2017 validation images~\cite{lin2014microsoft} with a fixed seed. All methods share the same per-prompt generation seed and initial noise, enabling paired comparisons.

\subsection{Algorithms}
We sweep the number of denoising steps $N \in \{25, 50, 100\}$ under the DDIM schedulers. The parallel sliding-window size $P$ is fixed to $16$. Per-method convergence tolerances and solver hyperparameters are held fixed across step counts; each fixed-point iteration terminates once the per-step update residual falls below its tolerance $\tau$. For all ParaAnya variants, the cache tolerance is set to $\tau_{\mathrm{cache}} = 0.05$; its effect is analyzed in \cref{sec:cache_tol}.

\subsection{Implementation Details}
The main comparison (\Cref{tab:main}) and the GPU-scaling experiment (\Cref{tab:num_gpus}) are conducted on a single node equipped with eight NVIDIA Tesla V100-SXM2 32GB GPUs and 16 CPU cores. The window-size and cache-tolerance ablations (\Cref{tab:window_size,tab:cache_tol}) are conducted on a single node equipped with eight NVIDIA H100 80GB HBM3 GPUs.
Every parallel method uses all eight GPUs and the serial baseline uses one GPU.
Before the timed loop, each method performs one untimed warm-up generation to exclude one-time CUDA and kernel setup costs.

\section{Results}
\label{sec:results}
\Cref{tab:main} reports the main results, with the serial single-GPU sampler as the $1.00\times$ reference.

\textbf{The cache matches or improves wall-clock speedup in every configuration.}
The proposed cache mechanism (ParaAnya) consistently improves wall-clock speedups across all evaluated configurations. As shown in the \Cref{tab:main}, integrating the cache into existing baseline methods, including ParaDiGMS, ParaSolver, ParaTAA, and SRDS, increases the speedup for every tested DDIM step count. Notably, ParaAnya achieves up to a 5.62$\times$ wall-clock speedup when integrated with ParaTAA. This universal improvement is driven by a reduction in the Number of Function Evaluations per image (NFE/img) in the ParaAnya rows, demonstrating that the caching mechanism successfully eliminates redundant computations and delivers consistent wall-clock speedups.

\textbf{Quality is preserved.} 
These performance gains come with negligible changes in CLIP scores across all evaluated configurations, suggesting that caching preserves text-image alignment while accelerating sampling. The differences are particularly small for SRDS, remaining effectively zero across all step counts.

\subsection{The Effect of Window Size}
\label{subsec:window}
We evaluate how the window size ($P$) affect the efficiency of ParaSolver with and without ParaAnya on 8 GPUs, using DDPM ($T=1000$) and DDIM ($T=50$).
\cref{tab:window_size} shows that increasing the window size does not always improve speedup, reflecting a trade-off between parallelism and convergence cost. Small windows limit the work available for parallel execution, potentially leaving GPU throughput underutilized. Conversely, overly large windows extend the prediction horizon of Picard iteration and may require more iterations to converge, offsetting the benefit of increased parallelism. For DDPM, both methods achieve their highest speedup at $P=64$ and slow down at $P=128$. For DDIM, the best window size is $P=8$ for ParaSolver and $P=16$ for ParaAnya, with both showing lower speedups at $P=32$. These results highlight the importance of choosing a window size that balances parallel execution with convergence efficiency.

\begin{table}[htbp]
\centering
\caption{Window-size speedup sweep on COCO2017 (1,000 captions) using 8 NVIDIA H100 80GB HBM3 GPUs. The 1-GPU serial baselines are 21.920\,s/img (DDPM) and 1.065\,s/img (DDIM). \textbf{Bold} marks the best speedup per row; \underline{underlining} marks ParaAnya speedups above uncached ParaSolver at the same $P$.}
\label{tab:window_size}
\renewcommand{\arraystretch}{1.0}
\setlength{\tabcolsep}{2pt}
\footnotesize
\begin{tabular}{@{}lccccc@{}}
\toprule
\multicolumn{6}{c}{\textbf{DDPM, $T=1000$}} \\
\midrule
Method & $P=8$ & $P=16$ & $P=32$ & $P=64$ & $P=128$ \\
\midrule
ParaSolver & 4.46$\times$ & 7.63$\times$ & 11.05$\times$ & \textbf{12.41$\times$} & 9.76$\times$ \\
ParaSolver+ParaAnya & \underline{4.49$\times$} & \underline{7.69$\times$} & \underline{11.09$\times$} & \textbf{\underline{13.76$\times$}} & \underline{13.38$\times$} \\
\midrule
\multicolumn{6}{c}{\textbf{DDIM, $T=50$}} \\
\midrule
Method & $P=8$ & $P=16$ & $P=32$ & -- & -- \\
\midrule
ParaSolver & \textbf{1.98$\times$} & 1.93$\times$ & 1.52$\times$ & -- & -- \\
ParaSolver+ParaAnya & \underline{2.80$\times$} & \textbf{\underline{3.08$\times$}} & \underline{2.60$\times$} & -- & -- \\
\bottomrule
\end{tabular}
\end{table}

\subsection{The Effect of Cache Tolerance}
\label{sec:cache_tol}
We investigate how the cache tolerance $\tau_{\mathrm{cache}}$ affects the efficiency and generation quality of ParaSolver on COCO2017. A larger tolerance permits more cached results to be reused, reducing fresh model evaluations but potentially introducing greater approximation error.
Table~\ref{tab:cache_tol} reports iteration counts, NFE, cache hits, average runtime, CLIP score, and speedup. All speedups are measured against the same 1-GPU serial baseline, which requires 1.065\,s per image with NFE 50.

Increasing $\tau_{\mathrm{cache}}$ consistently increases cache hits and reduces both iteration counts and NFE across the tested settings. However, these reductions do not always translate into lower runtime at small tolerances. At $\tau_{\mathrm{cache}}=0.001$ and $0.005$, cache reuse is limited, and runtimes of 0.579 and 0.582\,s per image are slightly higher than the uncached runtime of 0.551\,s. Caching first improves runtime over uncached ParaSolver at $\tau_{\mathrm{cache}}=0.01$, achieving a $2.01\times$ speedup. Larger tolerances yield more substantial gains: at $\tau_{\mathrm{cache}}=0.1$, NFE decreases from 176 to 57 and runtime falls to 0.299\,s per image, giving the highest speedup in this sweep, $3.56\times$. The reduction in iteration count becomes smaller between $\tau_{\mathrm{cache}}=0.05$ and $0.1$ (9.0 to 8.8), although NFE and runtime continue to decrease.

CLIP scores remain close to the uncached value throughout the sweep. The score is unchanged at 31.26 for tolerances up to 0.01 and reaches 31.29 at 0.05, before decreasing slightly to 31.19 at 0.1. Thus, cache tolerance provides a tunable trade-off between computational efficiency and generation quality as measured by CLIP. Among the tested settings, $\tau_{\mathrm{cache}}=0.05$ offers a favorable balance, achieving a $3.08\times$ speedup without a decrease in CLIP score, while $\tau_{\mathrm{cache}}=0.1$ provides the fastest generation with a small reduction in this quality metric.

\begin{table}[t]
\centering
\caption{Cache-tolerance sweep on COCO2017 using 8 NVIDIA H100 80GB HBM3 GPUs. ``None'' is uncached ParaSolver. Speedup uses this sweep's 1-GPU serial baseline (1.065\,s/img, NFE 50). Bold marks the best speedup; underlining marks speedups above uncached ParaSolver.}
\label{tab:cache_tol}
\renewcommand{\arraystretch}{1.0}
\small
\begin{tabular}{@{}rrrrrrr@{}}
\toprule
$\tau_{\mathrm{cache}}$ & Iters & NFE & Hits & Time & CLIP$\uparrow$ & Speedup \\
\midrule
None & 14.2 & 176 & --- & 0.551 & 31.26 & 1.93$\times$ \\
\midrule
0.001 & 14.2 & 175 & 0.2 & 0.579 & 31.26 & 1.84$\times$ \\
0.005 & 13.6 & 169 & 3.9 & 0.582 & 31.26 & 1.83$\times$ \\
0.01 & 12.5 & 152 & 11.5 & 0.529 & 31.26 & \underline{2.01$\times$} \\
0.05 & 9.0 & 70 & 41.1 & 0.346 & 31.29 & \underline{3.08$\times$} \\
0.1 & 8.8 & 57 & 52.1 & 0.299 & 31.19 & \textbf{\underline{3.56$\times$}} \\
\bottomrule
\end{tabular}
\end{table}

\subsection{The effect of the number of GPUs}
\label{subsec:gpus}
Table \ref{tab:num_gpus} evaluates the scaling behavior of various parallelization methods on DDIM ($T=50$) across 1, 2, 4, and 8 GPUs. The empirical results establish two primary conclusions regarding hardware scalability and the algorithmic efficiency of ParaAnya.
First, the generation speedup increases with the number of GPUs for all evaluated methods. For example, the base ParaTAA method scales from a 0.49$\times$ speedup on a single GPU to 2.31$\times$ on 8 GPUs. While several base methods, including ParaDiGMS and ParaSolver, incur severe overhead on smaller GPU clusters (yielding sub-1.0$\times$ performance on 1 or 2 GPUs), scaling to 8 GPUs results in net-positive acceleration across all baselines.
Second, integrating ParaAnya improves performance across all evaluated base methods and GPU configurations, as shown by the bolded results in Table \ref{tab:num_gpus}. On 8-GPUs, ParaAnya increases the speedup of ParaDiGMS from 1.25$\times$ to 3.13$\times$, ParaSolver from 1.90$\times$ to 3.70$\times$, and ParaTAA from 2.31$\times$ to 3.79$\times$. These benefits also extend to configurations with fewer GPUs, where the base methods often underperform the reference baseline. For instance, on a single GPU, ParaAnya improves ParaSolver from 0.41$\times$ to 1.02$\times$, turning a substantial slowdown into a modest speedup. Together, these results demonstrate that ParaAnya provides consistent performance improvements across the tested hardware configurations.

\begin{table}[t]
\centering
\caption{Speedup vs. the number of NVIDIA Tesla V100-SXM2 32GB GPUs for DDIM ($T\!=\!50$, $P\!=\!16$) on 1,000 COCO2017 captions. Baseline is 1-GPU serial generation (2.658\,s/img). ParaAnya results that outperform the corresponding base method are in \textbf{bold}.}
\label{tab:num_gpus}
\renewcommand{\arraystretch}{1.0}
\setlength{\tabcolsep}{4pt}
\small
\begin{tabular}{lcccc}
\toprule
Method & 1 GPU & 2 GPUs & 4 GPUs & 8 GPUs \\
\midrule
ParaDiGMS & 0.26$\times$ & 0.48$\times$ & 0.83$\times$ & 1.25$\times$ \\
ParaDiGMS+ParaAnya & \textbf{0.83$\times$} & \textbf{1.42$\times$} & \textbf{2.14$\times$} & \textbf{3.13$\times$} \\
\midrule
ParaSolver & 0.41$\times$ & 0.74$\times$ & 1.28$\times$ & 1.90$\times$ \\
ParaSolver+ParaAnya & \textbf{1.02$\times$} & \textbf{1.71$\times$} & \textbf{2.75$\times$} & \textbf{3.70$\times$} \\
\midrule
ParaTAA & 0.49$\times$ & 0.90$\times$ & 1.57$\times$ & 2.31$\times$ \\
ParaTAA+ParaAnya & \textbf{0.92$\times$} & \textbf{1.60$\times$} & \textbf{2.73$\times$} & \textbf{3.79$\times$} \\
\midrule
SRDS & 0.99$\times$ & 1.45$\times$ & 1.79$\times$ & 2.26$\times$ \\
SRDS+ParaAnya & \textbf{1.24$\times$} & \textbf{1.91$\times$} & \textbf{2.38$\times$} & \textbf{3.19$\times$} \\
\bottomrule
\end{tabular}
\end{table}

\section{Conclusion}
We introduced ParaAnya, an output cache that is agnostic to the parallel sampling algorithm.
It reuses cached network outputs when the current input latent is sufficiently close to the corresponding cached input, without modifying the underlying algorithm’s update rule. 
In the evaluated settings, the cached iteration produces solutions close to those of the uncached baseline.
Integrated into four parallel-in-time algorithms, ParaAnya matches or improves wall-clock speedup in \emph{every} tested configuration, reaching $5.62\times$ on 8 GPUs while maintaining CLIP scores comparable to the uncached baselines. 
ParaAnya delivers substantial speedups over the serial baseline when paired with parallel sampling algorithms on eight GPUs.
These results show that the excess network evaluations in parallel-in-time sampling are not solely an unavoidable cost of parallelism: a substantial portion is redundancy that a systems-level cache can eliminate while preserving the underlying update rules. Natural extensions include online adaptation of $\tau_{\mathrm{cache}}$, support for higher-order solvers, and applications to video and audio diffusion.

\bibliographystyle{IEEE}
\bibliography{refs}

\end{document}